\documentclass[prd,aps,preprint,nofootinbib,superscriptaddress,tightenlines]{revtex4}
\usepackage{epsfig}
\usepackage{amsmath}

\begin{document}


\preprint{RBRC-612} \preprint{SLAC-PUB-12167}

\title{Single Transverse-Spin Asymmetries at Large-x}

\author{Stanley J. Brodsky}
\affiliation{Stanford Linear Accelerator Center, Stanford
University, Stanford, California 94309}
\author{Feng Yuan}
\affiliation{RIKEN/BNL Research Center, Building 510A, Brookhaven
National Laboratory, Upton, NY 11973} \affiliation{Beijing
Institute of Modern Physics, Beijing University, Beijing 100871,
People's Republic of China}

\begin{abstract}
The large-$x$ behavior of the transverse-momentum dependent quark
distributions is analyzed in the factorization-inspired
perturbative QCD framework, particularly for the naive
time-reversal-odd quark Sivers function which is responsible for
the single transverse-spin asymmetries in various semi-inclusive
hard processes. By examining the dominant hard gluon exchange
Feynman diagrams, and using the resulting power counting rule, we
find that the Sivers function has power behavior  $(1-x)^4$ at $x
\to 1$, which is one power of $(1-x)$ suppressed relative to the
unpolarized quark distribution. These power-counting results
provide important guidelines for the parameterization of  quark
distributions and quark-gluon correlations.
\end{abstract}

\maketitle

\newcommand{\be}{\begin{equation}}
\newcommand{\ee}{\end{equation}}
\newcommand{\ben}{\[}
\newcommand{\een}{\]}
\newcommand{\beqn}{\begin{eqnarray}}
\newcommand{\eeqn}{\end{eqnarray}}
\newcommand{\Tr}{{\rm Tr} }
\vskip 2cm

\section{Introduction}

Single transverse-spin asymmetries (SSA) have a long history,
starting from the observation of large SSAs in hadron production
in nucleon-nucleon scattering in late 70s and 80s \cite{fixed}.
Since initial or final state phases are required to produce these
$T-odd$ observables, SSAs provide a unique window into quantum
chromodynamics (QCD) at the amplitude level as well as  the role of quark
orbital angular momentum in the wave functions of hadrons.
Recent experimental observations of sizeable SSAs in hard
scattering reactions such as single-inclusive deep inelastic
scattering  $\ell p \to \ell^\prime \pi^+ X$ have greatly
motivated new studies of the underlying mechanisms in QCD. These
experiments include semi-inclusive deep inelastic scattering
at HERMES at DESY \cite{hermes}, COMPASS at CERN \cite{compass},
and Jlab \cite{jlab}, and  hadron production in nucleon-nucleon
scattering at RHIC \cite{star,phenix,brahms}. Large SSAs have been
observed in semi-inclusive production of hadrons in DIS for
transversely polarized proton target at HERMES \cite{hermes}, and
single inclusive hadron production in the forward direction in
polarized proton-proton scattering at RHIC \cite{star,brahms}.

On the theory side, two mechanisms have been proposed in the QCD
framework to explain these large SSAs in hard scattering
processes. One is based on the QCD collinear factorization where
the asymmetries arise from the higher-twist quark-gluon
correlation effects (Efremov-Teryev-Qiu-Sterman mechanism)
\cite{et,qs}. Another approach explicitly takes into account the
effects coming from the intrinsic transverse momentum of partons
in hadrons. For example, the Sivers function was proposed in
\cite{Siv90} to explain the SSA phenomena in hadronic reactions,
where intrinsic transverse momentum plays an important role.

In the last few years, there has been an intensive theoretical
development of transverse momentum dependent (TMD) parton
distributions and their roles in  semi-inclusive processes such as
semi-inclusive deep inelastic scattering (SIDIS) and the small
transverse momentum Drell-Yan process. The gauge-invariant
properties\cite {bhs,col,bjy,bmp} of the TMD parton distributions
and the relevant factorization formalism
\cite{colsop81,css,jmy04,colmet} have been studied thoroughly. For
example, the Sivers effect in SIDIS has been shown to arise from
the interference of amplitudes differing by one unit of quark
orbital angular momentum and the fact that these amplitudes have
different final-state phases \cite{bhs}. The phases arise from the
Wilson-line associated with the struck quark as required by gauge
invariance \cite{col}. The SSA reverses sign in Drell-Yan
reactions because the phases in the Drell-Yan reaction arises from
initial-state rather than final-state interactions \cite{col,bhs}.
Remarkably, the SSA effect in these reactions is leading twist;
i.e., it survives in the Bjorken-scaling limit. Moreover, it was
recently shown that the above two mechanisms for SSAs are unified
for physical processes in the kinematical region where both apply
\cite{jqvy}.

There has also been a number of phenomenological studies of the
experimental data. Model-dependent parameterizations of the
relevant non-perturbative parton distributions (twist-3
quark-gluon correlation or the TMD quark distributions) have been
adopted to fit to the data \cite{qs,ans,met,vogyua,kqvy}. In these
studies it has been implicitly assumed the Sivers function
is suppressed at large $x$ relative to the unpolarized
quark distributions \cite{met,vogyua}. In this
paper, we will provide an argument for this suppression based on
power-counting of the leading diagrams in perturbative QCD. We
will utilize the generalized power counting rule and adopt a
perturbative analysis of the structure function at large $x.$

The large-x behavior of both the polarized and unpolarized parton
distributions have been studied
\cite{jackson,brolep,mul,gunion,bbs} in PQCD.  A generic factorization
has recently been used to justify the power counting rule by relating
parton distributions at large-$x$ to the
quark distribution amplitudes of hadrons \cite{fac}. So far, the
power counting results have been worked out for the unpolarized
and longitudinal polarized quark distributions. In the present
study, we will extend this analysis to other leading-order TMD quark
distributions, including the naive time-reversal-odd quark Sivers
function which is responsible for the SSAs in various
semi-inclusive hard processes.

It is important to note that the $x \to 1$ regime where the struck
quark has nearly all of the light-cone momentum of its parent
hadron involves dynamics far-off the mass shell: the  Feynman
virtuality of the struck quark becomes highly space-like: $k^2_F -
m^2 \sim -{k^2_\perp + {\cal M}^2\over 1-x}$, where $k_\perp$ and
$\cal M$ are the transverse momentum and invariant mass of the
spectator system. Thus we can use perturbative QCD to analyze the
large-$x$ behavior of parton distributions since the internal
propagators in the relevant Feynman diagrams scale as $1/(1-x)$.
This behavior leads to a power counting rule. This is because more
partons in hadron's wave function means more propagators in the
scattering amplitudes, and more suppression for the contribution
to the parton distributions. Thus the parton distributions at
large-$x$ depend on the number of partons in the Fock state wave
function of the hadron. In particular, the valence  Fock state
with the minimum number of constituents will dominate the quark
distribution function at large-$x$. For example, the proton
structure function will be dominated by its three-quark Fock
states, which can be further classified according to its quark
orbital angular momentum projection: $L_z=0$, $|L_z|=1$, or
$|L_z|=2$ \cite{jmy03}. Since nonzero quark orbital angular
momentum light-cone wave function normally introduces additional
suppression of $(1-x)$, we will consider in this paper only
$L_z=0$ and $|L_z|=1$ Fock states contributions. The $|L_z|=1$
state is needed because some of the TMD quark distributions
involve the interference between $L_z=0$ and $L_z=1$ states
\cite{bhs,{Brodsky:2006ha},jmy03} (see also the discussions
below).

As is the case of the nucleon form factors (Dirac and Pauli form
factors) \cite{stan-drell}, the transverse momentum dependent
quark distributions can be calculated from the overlap of the
light-cone wave functions of three-quark Fock states
\cite{bhs,{Brodsky:2006ha},jmy03}. As we shall demonstrate, the
large-x power counting for the TMD parton distributions can be
obtained in a similar manner. For example, we know that the
unpolarized quark distribution has power counting of $(1-x)^3$ at
large-x \cite{jackson}, which comes from the quark orbital angular
momentum projection $L_z=0$ Fock states contribution, whereas the
contribution from the overlap of two light-cone functions for
$|L_z|=1$ states is suppressed by
$(1-x)^2$\cite{brolep,mul,gunion,bbs}. On the other hand, since
the Sivers function depends on the interference between $L_z=0$
and $|L_z|=1$ states, simple counting suggests that that the
Sivers function will have the leading power of $(1-x)^4$. The
detailed calculations in this paper support this intuitive
argument.

The remainder of the paper is organized as follows. In Sec. II, we
present our analysis of the leading-order TMD quark distributions
at large-x, where we discuss the power counting results for the
$k_\perp$-even, $k_\perp$-odd, and naive time-reversal-odd quark
distributions respectively. We will also derive the power counting
results for the integrated parton distributions at leading-twist
and sub-leading-twist. We summarize our results in Sec.III.

\section{Transverse-momentum dependent quark
distributions at large-x}

The TMD quark distributions can be defined through the following
matrix:
\begin{eqnarray}
    {\cal M}^{\alpha\beta} &=&   P^+\int
        \frac{d\xi^-d^2\xi_\perp}{(2\pi)^3}e^{-ix\xi^-P^++i\xi_\perp\cdot k_\perp} \, \left\langle
PS\left|\overline{\mit \Psi}_v^\beta(\xi)
        {\mit \Psi}_v^\alpha(0)\right|PS\right\rangle\ ,
\end{eqnarray}
where the vector $P=(P^+,0^-,0_\perp)$ is along the momentum
direction of the proton, $S$ is the polarization vector, and
${\mit \Psi}_v(\xi)$ is defined as
\begin{equation}
{\mit \Psi}_v(\xi) \equiv {\cal L}_{v}(\infty;\xi)\psi(\xi)\ ,
\end{equation}
with the gauge link $ {\cal L}_{v}(\infty;\xi) \equiv
\exp\left(-ig\int^{\infty}_0 d\lambda \, v\cdot A(\lambda v
+\xi)\right)$. In this paper, we will study the TMD quark
distributions for the semi-inclusive DIS processes, thus the above
gauge link goes to $+\infty$. Our results can be simply extended
to the TMD quark distributions for the Drell-Yan process, where
the naive time-reversal-odd TMDs will have an opposite sign
\cite{bhs,col,bjy}. In the following analysis, no light-cone
singularity will be present, thus we can choose the vector $v$ to
be a light-cone vector $n=(0^+,n^-,0_\perp)$ with $n\cdot P=1$.

The leading order expansion of the matrix ${\cal M}$ contains
eight quark distributions \cite{mulders98}. Among the eight
distributions, three are the so-called $k_\perp$-even TMD quark
distributions: $q(x,k_\perp)$, $\Delta q_L(x,k_\perp)$, and
$\delta q_T(x,k_\perp)$, which correspond to the unpolarized,
longitudinal polarized, and transversity distributions,
respectively. These distributions will lead to the three
leading-twist integrated quark distributions \cite{JafJi} after
integrating over transverse momentum. The other five distributions
are $k_\perp$-odd, and vanish when $k_\perp$ are integrated. Two
of them, $q_T(x,k_\perp)$ (Sivers) and $\delta q(x,k_\perp)$
(Boer-Mulders), are odd under naive time-reversal transformation
\cite{Col93}. The notations for these distributions follow Ref.
\cite{jmy03}, which are different than those in \cite{mulders98}.
However, their definitions are identical.

\begin{figure}[]
\begin{center}
\includegraphics[height=3.0cm]{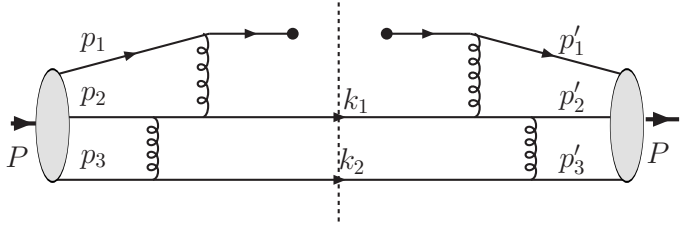}
\end{center}
\vskip -0.7cm \caption{\it Typical Feynman diagram contributing to
the large-x quark distribution in nucleon. The blobs at the left
and right sides represent the three-quark light-cone wave function
distribution amplitudes of the nucleon.}
\end{figure}

In this paper we are interested in studying the large-x behavior
of these TMD quark distributions. We will disregard the $k_\perp$
dependence, and further choose $k_\perp\gg \Lambda_{\rm QCD}$ in
order to avoid the infrared divergence associated with low
transverse momentum limit. A typical Feynman diagram contributing
to large-x quark distributions is shown in Fig.~1. At this order,
we can write down an inspired factorization formula for the the
parton distributions in terms of the distribution amplitudes of
the nucleon \cite{brolep,mul},
\begin{eqnarray}
f(x,k_\perp)&=&\int\frac{d^2k_{1\perp}d^2k_{2\perp}}{4(2\pi)^6}\frac{dz_1dz_2}{z_1z_2}
\delta(k_\perp+k_{1\perp}+k_{2\perp})\delta(z_1+z_2+x-1)
\int[dy_i][dy_i'] \nonumber\\
&&\Phi(y_i)\Phi'(y_i'){\cal H}(y_i,y_i';k_{i\perp};z_i)\ ,
\label{e5}
\end{eqnarray}
where the outside integral represents the phase space integrals
for the final state two quarks going through the cut line, with
momenta: $k_i=(z_iP^+,k_i^-,k_{i\perp})$ ($i=1,2$). The inside
integral measure $[dy_i]$ is defined as
$[dy_i]=dy_1dy_2dy_3\delta(1-y_1-y_2-y_3)$, and the $y_i$ are the
momentum fractions of the proton carried by the quarks in the
light cone wave functions, i.e., $p_i=y_iP$ and $p_i'=y_i'P$ in
Fig.~1. Here $f$ represents any of the leading order TMD quark
distributions. $\Phi$ and $\Phi'$ represent the quark distribution
amplitudes of nucleon at the left and right sides of the cut line,
respectively. They can be the leading-twist or higher-twist
distribution amplitudes, depending on the quark orbital angular
momentum projection along $z$-direction. We list the distribution
amplitudes we will use in this paper in the Appendix for
reference. ${\cal H}$ represents the hard part which can be
calculated from the perturbative Feynman diagram like Fig.~1.

We notice that, in the above equation the phase space integral for
$k_1$ and $k_2$ are strongly constrained in the limit of $x \to
1$, because of momentum conservation, $z_1+z_2=1-x$. We can factor
out the $(1-x)$ dependence of this phase space integral, taking
the following parameterizations: $z_1=\alpha
(1-x),~z_2=\beta(1-x)$,
\begin{equation}
\int \frac{dz_1dz_2}{z_1z_2}\delta(z_1+z_2+x-1)=\frac{1}{1-x}\int
\frac{d\alpha d\beta}{\alpha\beta}\delta(1-\alpha-\beta) \ .
\label{phase}
\end{equation}
This leads to an overall enhancement of $1/(1-x)$. After factoring
this out, the remaining measure of the phase space integral ($d\alpha
d\beta$) does not contain any additional factors of $(1-x)$.

Additional $(1-x)$ factors can come from the hard amplitude ${\cal
H}$, but these depend on the structure of the relevant tree
diagrams. Since the hard propagators each contain a $(1-x)$
factor, the least number of active particles involved in the hard
process lead to the least suppression. Thus the leading
contribution to the quark distributions at large-x is dominated by
the leading component in the hadron's Fock state expansion. For
the nucleon, the three-quark Fock state components will dominate
the quark distributions, while for pion it will be the
quark-antiquark pair states. In the following we will study the
large-x power counting for the above mentioned TMD quark
distributions, including the three $k_\perp$-even ones: $q$,
$\Delta q_L$, $\delta q_T$, and four $k_\perp$-odd ones: $\delta
q_L$, $\Delta q_T$, $q_T$, and $\delta q$. For $\delta q_T'$, its
analysis will involve much more complicated diagrams, and we will
not discussed this in the present paper.

\subsection{$k_\perp$-even quark distributions}

The unpolarized quark distribution is defined as
\begin{equation}
q(x,k_\perp)=\frac{1}{2}\int
        \frac{d\xi^-d^2\xi_\perp}{(2\pi)^3}e^{-ix\xi^-P^++i\xi_\perp\cdot k_\perp} \left\langle
P\left|\overline{\mit \Psi}_v(\xi)\gamma^+
        {\mit \Psi}_v(0)\right|P\right\rangle\ .
\end{equation}
The large-x power counting for this distribution function has been
studied in the literature \cite{jackson,brolep}. In the following,
we will repeat these arguments as guideline for the analysis of
other quark distributions.

We calculate the above matrix element in the proton helicity
basis,
\begin{equation}
q(x,k_\perp)=\frac{1}{2}\int
        \frac{d\xi^-d^2\xi_\perp}{(2\pi)^3}e^{-ix\xi^-P^++i\xi_\perp\cdot k_\perp} \frac{1}{2}\left(
        \langle
PS_{z\uparrow}|\hat{\cal O}|PS_{z\uparrow}\rangle+\langle
PS_{z\downarrow}|\hat{\cal O}|PS_{z\downarrow}\rangle\right)\ ,
\label{eq}
\end{equation}
where the operator $\hat {\cal O}$ is defined as $\hat{\cal
O}=\overline{\mit \Psi}_v(\xi)\gamma^+{\mit \Psi}_v(0)$. This
operator is chiral-even, and conserves the quark helicity in the
partonic scattering matrix elements, and so that the dominant
contributions come from the leading Fock state wave function
($L_z=0$) at both sides of the cut in Fig.~1. The matrix element
$\langle PS_{z\uparrow}|\hat{\cal O}|PS_{z\uparrow}\rangle$ will
have the contributions from the following quark spin
configurations: $\uparrow\downarrow\uparrow$ and
$\downarrow\uparrow\uparrow$, where in the first one the probing
quark has the same helicity as the proton and in the second case
it is opposite. If the probing quark's spin is parallel to the
proton spin, the two spectator quarks will form a scalar.
\begin{figure}[]
\begin{center}
\includegraphics[height=4.5cm]{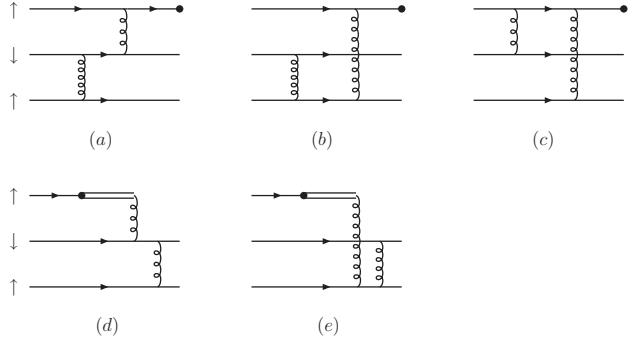}
\end{center}
\vskip -0.7cm \caption{\it Leading diagrams contributing to the
$k_\perp$-even quark distributions at large-x: left half of the
relevant diagrams are shown. The contributions will the amplitudes
square of these diagrams, including the interference between them.
These diagrams also contribute to the $k_\perp$-odd and naive
time-reversal-odd TMD quark distributions.}
\end{figure}

In the following, we are only interested in obtaining the power
counting for the quark distributions, and the explicit dependence
on the distribution amplitudes will not be discussed. According to
the factorization formula Eq.~(\ref{e5}) and the reduced integral
Eq.~(\ref{phase}), the unpolarized quark distribution depends on
the leading-twist distribution amplitudes of the nucleon,
\begin{equation}
q(x,k_\perp)|_{x\to 1}\propto \frac{1}{1-x}\int\frac{d\alpha
d\beta}{\alpha\beta} \Phi_3(y_i)\Phi_3(y_i'){\cal
H}\left(y_i;y_i';\alpha,\beta,(1-x)\right) \ .
\end{equation}
The power counting of the hard factor ${\cal H}$ can be evaluated
from the partonic scattering matrix elements, which include a set
of propagators and traces of the Dirac matrices. As mentioned
above, the propagators are far off-shell in the limit of $x\to 1$,
which will lead to suppression in terms of $(1-x)$. For example,
One of the gluon propagator in Fig.~1 goes like,
\begin{equation}
\frac{1}{(p_3-k_2)^2}=\frac{1}{2p_3\cdot k_2}\approx
-\frac{1}{\langle k_\perp^2\rangle}\frac{1-x}{y_3} \ ,
\end{equation}
at large $x$. In the above expression, we have omitted all higher
order terms suppressed by $(1-x)$. $\langle k_\perp^2\rangle$
represents a typical momentum scale in order of transverse
momentum $k_\perp$. Besides the propagators, the traces of the
Dirac matrix contains $(1-x)$ factors as well, which will depend
on the spin structure of the quarks in the scattering matrix
elements. For example, the traces of the Dirac matrices for the
scattering $\uparrow\downarrow\uparrow\to
\uparrow\downarrow\uparrow$ in  Fig.~1 will contribute to the
matrix element $\langle PS_{z\uparrow}|\hat{\cal
O}|PS_{z\uparrow}\rangle$ as
\begin{equation}
\frac{1}{(1-x)^4} \ ,\label{trace}
\end{equation}
in the leading power, where the probing quark has the same
helicity as the proton. If the probing quark has an opposite
helicity as the proton, e.g., in the spin structure
$\downarrow\uparrow\uparrow$, the Dirac trace for the diagram of
Fig.~1 vanishes. We have checked all other diagrams, and found
that the Dirac traces for those diagrams with spin-one
configuration for the spectator quarks ($\uparrow\uparrow$ or
$\downarrow\downarrow$ for the two spectator quarks) either vanish
or are suppressed by at least $(1-x)^2$ as compared to the scalar
configuration ($\uparrow\downarrow$ or $\downarrow\uparrow$). This
property has long been noticed in the literature \cite{jackson,
brolep,gunion}. In Fig.~2, we showed all leading diagrams for the
spin structure of $\uparrow\downarrow\uparrow$, where only left
half sides of the relevant diagrams are shown. The contributions
will be the amplitudes square of these diagrams, including their
interferences.

In summary, the leading contribution to the matrix element in
Eq.~(\ref{eq}) comes from the quark spin structure with probing
quark's helicity equal to the nucleon's helicity. If they differ,
the contribution will be suppressed by $(1-x)^2$. This is to say,
the quark distribution is dominated by the quark spin parallel to
the nucleon spin, and the quark spin anti-parallel distribution
will be suppressed by $(1-x)^2$.

The final power counting result will depend on the above factors,
including the Dirac matrix traces, the power counting of the
propagators, and the phase spaces integrals. For example, the
diagram in Fig.~1 contains eight propagators, with overall power
counting,
\begin{equation}
\sim \frac{(1-x)^8}{y_3(1-y_1)^2y_3'(1-y_1')^2} \ .
\end{equation}
Combining the above with the contribution from the Dirac matrix
traces in Eq.~(\ref{trace}) and the phase spaces integral factor
in Eq.~(\ref{phase}), we find that the contribution of this
diagram to the unpolarized quark distribution is
\begin{equation}
q(x,k_\perp)|_{x\to 1}\sim (1-x)^3 \ .
\end{equation}
All other diagrams in Fig.~2 will contribute the same power to the
unpolarized quark distribution. For example, the amplitude square
of Fig.~2(d) contribute a power of $(1-x)^{-2}$ from the Dirac
matrix traces, a power of $(1-x)^6$ from the propagators, plus a
power of $(1-x)^{-1}$ from the phase space integral
Eq.~(\ref{phase}), which leads to a power of $(1-x)^3$
contribution to the quark distribution.

The longitudinal polarized quark distribution can be analyzed
accordingly, which is defined through the following matrix
element,
\begin{equation}
\Delta q_L(x,k_\perp)=\frac{1}{2}\int
        \frac{d\xi^-d^2\xi_\perp}{(2\pi)^3}e^{-ix\xi^-P^++i\xi_\perp\cdot k_\perp} \left\langle
PS_z\left|\overline{\mit \Psi}_v(\xi)\gamma^+\gamma_5
        {\mit \Psi}_v(0)\right|PS_z\right\rangle\ .
\end{equation}
Again, if we calculate in the proton helicity states, we will
obtain
\begin{equation}
\Delta q_L(x,k_\perp)=\frac{1}{2}\int
        \frac{d\xi^-d^2\xi_\perp}{(2\pi)^3}e^{-ix\xi^-P^++i\xi_\perp\cdot k_\perp} \frac{1}{2}\left(
        \langle
PS_{z\uparrow}|\hat {\cal O}_L|PS_{z\uparrow}\rangle- \langle
PS_{z\downarrow}|\hat {\cal O}_L|PS_{z\downarrow}\rangle\right)\ ,
\end{equation}
where the operator $\hat{\cal O}_L$ is defined as $\hat {\cal
O}_L=\overline{\mit \Psi}_v(\xi)\gamma^+\gamma_5 {\mit
\Psi}_v(0)$. We can interpret longitudinal polarized quark
distribution as the quark spin parallel to the nucleon spin
distribution minus the antiparallel distribution. According to the
above analysis for the unpolarized quark distribution, we know
that the quark spin parallel to the nucleon spin distribution
dominates over the antiparallel distribution, and the latter is
suppressed by an extra factor of $(1-x)^2$. In conclusion, we will
obtain the same power behavior for the longitudinal polarized
quark distribution as the unpolarized quark distribution,
\begin{equation}
\Delta q_L(x,k_\perp)|_{x\to 1}\sim (1-x)^3 \ .
\end{equation}

The transversity distribution for the quarks can be analyzed in
the same way. It is defined as
\begin{equation}
\delta q_T(x,k_\perp)=\frac{1}{2}\int
        \frac{d\xi^-d^2\xi_\perp}{(2\pi)^3}e^{-ix\xi^-P^++i\xi_\perp\cdot k_\perp}\langle
PS_\perp|\overline{\mit \Psi}_v(\xi)\gamma^+\gamma^\perp\gamma_5
        {\mit \Psi}_v(0)|PS_\perp\rangle\ .
\end{equation}
Here, the proton is transversely polarized. We can choose the
polarization vector along $x$-direction, and the polarization
states can be constructed from the proton helicity states,
\begin{eqnarray}
|PS_{x\uparrow}\rangle=\frac{1}{\sqrt{2}}\left(|PS_{{z\uparrow}}\rangle+
|PS_{z\downarrow}\rangle\right)\ ,~~~~
|PS_{x\downarrow}\rangle=\frac{1}{\sqrt{2}}\left(|PS_{z\uparrow}\rangle-
|PS_{z\downarrow}\rangle\right)\ .
\end{eqnarray}
Substituting the above into the definition of the transversity
distribution, we will obtain
\begin{equation}
\delta q_T(x,k_\perp)=\frac{1}{2}\int
        \frac{d\xi^-d^2\xi_\perp}{(2\pi)^3}e^{-ix\xi^-P^++i\xi_\perp\cdot k_\perp}\frac{1}{2}\left(
        \left\langle
PS_{z\uparrow}|\hat{\cal O}_t |PS_{z\downarrow}\right\rangle
+\left\langle PS_{z\downarrow}\left|\hat{\cal
O}_t\right|PS_{z\uparrow}\right\rangle\right)\ ,
\end{equation}
where the operator $\hat{\cal O}_t$ is defined as $\hat{\cal
O}_t=\overline{\mit
\Psi}_v(\xi^-,0,\vec{b}_\perp)\gamma^+\gamma^\perp\gamma_5 {\mit
\Psi}_v(0)$. From this equation, we see that the quark
transversity distribution depends on the matrix elements with
hadron helicity flip. On the other hand, because the operator
$\hat {\cal O}_t$ is chiral-odd, it changes the quark helicity in
the partonic scattering process as well. If we keep the leading
Fock state contribution, the matrix element in the bracket of the
above equation will reduce to
\begin{equation}
{}_{\frac{1}{2}} \left\langle PS_{z\uparrow}|\hat{\cal O}_t
|PS_{z\downarrow}\right\rangle {}_{-\frac{1}{2}}
+{}_{-\frac{1}{2}}\left\langle PS_{z\downarrow}\left|\hat{\cal
O}_t\right|PS_{z\uparrow}\right\rangle {}_{\frac{1}{2}}\ ,
\end{equation}
where the subscripts $\pm \frac{1}{2}$ represents the total quark
helicity in the three-quark wave function used in the
calculations. From this equation, we can easily see that it will
be the same set of diagrams in Fig.~2 contributing to the
transversity quark distributions. The same power counting results
will be obtained,
\begin{equation}
\delta q_T(x,k_\perp)|_{x\to 1}\sim (1-x)^3 \ .
\end{equation}
From the above analysis, all the three $k_\perp$-even quark
distributions have the same power behavior at large-x, which is
certainly consistent with the inequality condition for them
\cite{soffer}.

\subsection{$k_\perp$-odd and naive time-reversal-even quark
distributions}

In this subsection, we will study two $k_\perp$-odd but naive
time-reversal-even TMD quark distributions: $\Delta q_T$ and
$\delta q_T$, which represent the longitudinal polarized quark
distribution in a transversely polarized proton and the
transversely polarized quark distribution in a longitudinal
polarized proton, respectively. $\Delta q_T$ can be calculated
from the following matrix element,
\begin{equation}
\Delta q_T(x,k_\perp)=\frac{M_P}{2S_\perp\cdot k_\perp}\int
        \frac{d\xi^-d^2\xi_\perp}{(2\pi)^3}e^{-ix\xi^-P^++i\xi_\perp\cdot k_\perp}\, \left\langle
PS_\perp|\hat{\cal O}_L|PS_\perp\right\rangle\ ,
\end{equation}
where the operator $\hat {\cal O}_L$ as defined above. Following
the above calculation for the transversity distribution, we choose
the transverse polarization vector along the $x$-direction, and
the above equation can be reduced to
\begin{equation}
\Delta q_T(x,k_\perp)=\frac{M_P}{2k_\perp^x}  \int
        \frac{d\xi^-}{(2\pi)^3}e^{-ix\xi^-P^++i\xi_\perp\cdot k_\perp}
        \frac{1}{2}\left(
        \left\langle
PS_{z\uparrow}|\hat{\cal O}_L |PS_{z\downarrow}\right\rangle
+\left\langle PS_{z\downarrow}\left|\hat{\cal
O}_L\right|PS_{z\uparrow}\right\rangle\right)\ . \label{dqt}
\end{equation}
$\hat {\cal O}_L$ is a chiral-even operator, and it conserves the
quark helicity. On the other hand, the above matrix element has
hadron helicity flip, thus the total quark helicity and the hadron
helicity will mismatch on either side of the above matrix element.
If the total quark helicity and the proton helicity is
mismatching, the wave function for the three-quark state must have
nonzero quark orbital angular momentum. It is the interference
between the $L_z=0$ and $|L_z|=1$ states contributing to the TMD
quark distribution $\Delta q_T$.

In order to proceed, we further decompose the proton spin state
into the Fock states containing $L_z=0$ and $|L_z|=1$. For
example,
\begin{equation}
|PS_{z\uparrow}\rangle=|PS_{z\uparrow}\rangle_{1/2}+|PS_{z\uparrow}\rangle_{-1/2}
\ ,
\end{equation}
where the subscript $\pm 1/2$ denotes the total quark helicity.
The first term in the above equation represents the $L_z=0$ state,
while second one for the $L_z=1$ state. The wave function
parameterizations for these states have been given in
Eq.~(\ref{w1}). Similarly, for $|PS_{z\downarrow}\rangle$ we have,
\begin{equation}
|PS_{z\downarrow}\rangle=|PS_{z\downarrow}\rangle_{-1/2}+|PS_{z\downarrow}\rangle_{1/2}
\ ,
\end{equation}
where the first one is for $L_z=0$, and the second one for
$L_z=-1$. Because the partonic matrix element conserves the quark
helicity, in the calculation of the matrix element of
Eq.~(\ref{dqt}), the quarks helicities will remain the same at the
left and right sides of the cut line in the Feynman diagram like
Fig.~1. From the experience in the last subsection, we know that
the partonic processes where the two spectator quarks have
opposite helicities dominate the quark distributions at large-x.
So, for the leading contributions we will have two typical
partonic processes:
$\uparrow\downarrow\uparrow\to\uparrow\downarrow\uparrow$ with
total quark helicity $1/2$ and
$\downarrow\uparrow\downarrow\to\downarrow\uparrow\downarrow$ with
total quark helicity $-1/2$. These two actually will contribute
opposite sign to the matrix element in Eq.~(\ref{dqt}), because of
the $\gamma_5$ in the operator $\hat {\cal O}_L$. Taking into
account this fact, and substituting the above decomposition into
Eq.~(\ref{dqt}), we find that the matrix element becomes
\begin{eqnarray}
{}_{\frac{1}{2}}\langle PS_{z\uparrow}|\hat{\cal
O}_L|PS_{z\downarrow}\rangle_{\frac{1}{2}}&-&{}_{-\frac{1}{2}}\langle
PS_{z\downarrow}|\hat{\cal
O}_L|PS_{z\uparrow}\rangle_{-\frac{1}{2}}\nonumber\\
-{}_{-\frac{1}{2}}\langle PS_{z\uparrow}|\hat{\cal
O}_L|PS_{z\downarrow}\rangle_{-\frac{1}{2}}&+&{}_{\frac{1}{2}}\langle
PS_{z\downarrow}|\hat{\cal
O}_L|PS_{z\uparrow}\rangle_{\frac{1}{2}}\ . \label{shalf}
\end{eqnarray}
It is easy to see that the above two lines are complex conjugates.
In the following, we will consider the contribution from the first
line, and the other one can be obtained immediately.

For the subprocess
$\uparrow\downarrow\uparrow\to\uparrow\downarrow\uparrow$, the
contribution to the matrix element will be
\begin{equation}
{}_{\frac{1}{2}}\langle PS_{x\uparrow}|\hat {\cal
O}|PS_{x\downarrow}\rangle_{\frac{1}{2}} \propto \int
\tilde\psi^{(1)}(y_i')\tilde\psi^{(3)}(y_i,p_{i\perp})\left(p_1^x-ip_1^y\right)T_H(y_i;y_i';p_{i\perp})
\ , \label{e32}
\end{equation}
where $\tilde\psi^{(1)}$ is the wave function for $L_z=0$ Fock
state and $\tilde \psi^{(3)}$ for $|L_z|=1$ (their definitions are
listed in the Appendix). Here we only show the contribution from
the interference between $\tilde \psi^{(1)}$ and
$\tilde\psi^{(3)}$ wave functions, and other interference
contributions (e.g., the one with $\tilde\psi^{(1)}$ and
$\tilde\psi^{(4)}$) can be calculated similarly. Because proton is
stable, the light-cone wave functions are real, i.e., $(\tilde
\psi)^*=\tilde\psi$. Meanwhile, for the
$\downarrow\uparrow\downarrow\to \downarrow\uparrow\downarrow$
partonic process, we will have
\begin{equation}
{}_{-\frac{1}{2}}\langle PS_{x\downarrow}|\hat{\cal
O}|PS_{x\uparrow}\rangle{}_{-\frac{1}{2}} \propto \int
\tilde\psi^{(1)}(y_i')\tilde\psi^{(3)}(y_i,p_{i\perp})\left(-p_1^x-ip_1^y\right)T_H(y_i;y_i';p_{i\perp})
\ .\label{e33}
\end{equation}
The hard partonic parts $T_H$ in the above two equations are
identical to each other for the same diagram if we change all the
quarks helicities. Thus we can sum their contributions together,
and the matrix element will be
\begin{equation}
 \langle PS_{x\uparrow}|\hat {\cal
O}|PS_{x\downarrow}\rangle -\langle PS_{x\downarrow}|\hat {\cal
O}|PS_{x\uparrow}\rangle\propto \int
\tilde\psi^{(1)}(y_i')\tilde\psi^{(3)}(y_i,p_{i\perp})\left(p_1^x\right)T_H(y_i;y_i';p_{i\perp})
+h.c. \label{eg1t}
\end{equation}
The linear expansion term of $p_{i\perp}$ from $T_H$ will be
crucial to obtain nonzero contribution to the above matrix element
when integrating over $p_{i\perp}$. Otherwise, it will vanish.
This expansion will introduce an additional suppression factor in
$(1-x)$. For example, one of the propagators in Fig.~1 has the
following expansion result,
\begin{eqnarray}
\frac{1}{(p_3-k_2)^2}&=&\frac{1}{(y_3P-k_2+p_{3\perp})^2}\nonumber\\
&=&\frac{\beta(1-x)}{y_3k_{2\perp}^2}\left(1-
\frac{\beta(1-x)}{y_3k_{2\perp}^2}2p_{3\perp}\cdot
k_{2\perp}\right) \ . \label{e24}
\end{eqnarray}
Substituting the above into Eq.~(\ref{eg1t}), and using the fact
that $p_{3\perp}=-p_{1\perp}-p_{2\perp}$, we find that the above
expansion will lead to a contribution as $(1-x)\int d^2p_{1\perp}
p_{1\perp}^x \left(p_{1\perp}\cdot
k_{\perp}\right)\tilde\psi^{(3)}\sim (1-x)k_\perp^x \Phi^{(3,4)}$,
where $\Phi^{(34)}$ represents a combination of twist-four
distribution amplitudes $\Phi_4$ and $\Psi_4$, and the $k_\perp^x$
factor will cancel out the same $k_\perp^x$ in the denominator in
Eq.~(\ref{dqt}). This suppression feature applies to every
propagator expansion containing the linear term of the intrinsic
transverse momentum $p_\perp$. Similarly, the Dirac wave function
expansion in terms of $p_{\perp}$ will also be suppressed by
$(1-x)$.

The above analysis can be repeated for every diagrams in Fig.~2,
and they contribute the same. So, the final power counting result
for the TMD quark distribution $\Delta q_T$ will be,
\begin{equation}
\Delta q_T(x,k_\perp)|_{x\to 1}\sim (1-x)^4 \ .
\end{equation}

Similar analysis can be performed for the TMD quark distribution
$\delta q_L$, which is defined through the following matrix
element,
\begin{equation}
\delta q_L(x,k_\perp)=\frac{M_P}{2k_\perp^i}\int
        \frac{d\xi^-d^2\xi_\perp}{(2\pi)^3}e^{-ix\xi^-P^++i\xi_\perp\cdot k_\perp}\, \left\langle
PS_z|\overline{\mit \Psi}_v(\xi)\gamma^+\gamma^i\gamma_5 {\mit
\Psi}_v(0)|PS_z\right\rangle\ .
\end{equation}
If we choose $\gamma^i=\gamma^x$ in the above equation, the TMD
$\delta q_L$ will become,
\begin{equation}
\delta q_L(x,k_\perp)=\frac{M_P}{2k_\perp^x}\int
        \frac{d\xi^-d^2\xi_\perp}{(2\pi)^3}e^{-ix\xi^-P^++i\xi_\perp\cdot k_\perp} \langle
PS_z|\hat {\cal O}_t|PS_z\rangle\ ,
\end{equation}
where the operator $\hat{\cal O}_t$ follows the definition in the
subsection Sec.II(b). In the above definition, the proton is
longitudinal polarized, and we can further write down explicitly
in terms of the proton helicity states,
\begin{equation}
\delta q_L(x,k_\perp)=\frac{M_P}{2k_\perp^x}\int
        \frac{d\xi^-d^2\xi_\perp}{(2\pi)^3}e^{-ix\xi^-P^++i\xi_\perp\cdot k_\perp} \frac{1}{2}\left(\langle
PS_{z\uparrow}|\hat {\cal O}_t|PS_{z\uparrow}\rangle-\langle
PS_{z\downarrow}|\hat {\cal O}_t|PS_{z\downarrow}\rangle\right)\
.\label{h1l}
\end{equation}
Since the operator $\hat {\cal O}_t$ is chiral-odd, it changes the
quark helicity. However, in the above equation, we are calculating
the hadron helicity conserved matrix elements, thus the nonzero
quark orbital angular momentum projection must be taken into
account in order to obtain nonzero results, as in the case of
$\Delta q_T$ in the above. Following the above analysis, we find
the power counting result for the TMD quark distribution $\delta
q_L$,
\begin{equation}
\delta q_L|_{x\to 1}\sim (1-x)^4 \ ,
\end{equation}
which is one power of $(1-x)$ suppressed relative to the
unpolarized quark distribution.

\subsection{Naive time-reversal-odd quark distributions}

Now, we turn to the naive time-reversal-odd quark distributions.
At leading order, we have two: the quark Sivers function and the
Boer-Mulders function. The Sivers function represents the
unpolarized quark distribution in a transversely polarized target,
while the Boer-Mulders function represents the transversely
polarized quark distribution in a unpolarized proton target. These
two distributions are naive time-reversal-odd, and their existence
require final state interactions \cite{bhs}. The quark Sivers
function is defined as,
\begin{equation}
q_T(x,k_\perp)=\frac{M}{2\epsilon^{ij}S_\perp^ik_\perp^j}\int
        \frac{d\xi^-d^2\xi_\perp}{(2\pi)^3}e^{-ix\xi^-P^++i\xi_\perp\cdot k_\perp}
        \langle
PS|\hat {\cal O}|PS\rangle\ ,
\end{equation}
where the operator ${\cal O}$ follows the above definition.
Because the target is transversely polarized, again we will choose
the $x$-direction for its polarization, and the Sivers function
then becomes,
\begin{equation}
q_T(x,k_\perp)=\frac{M}{2k_\perp^y}\int\frac{d\xi^-d^2\xi_\perp}{(2\pi)^3}e^{-ik\cdot\xi}
\frac{1}{2}\left(\langle PS_{z\uparrow}|\hat{\cal
O}|PS_{z\downarrow}\rangle+\langle PS_{z\downarrow}|\hat{\cal
O}|PS_{z\uparrow}\rangle\right) \ .\label{e46}
\end{equation}
The above equation shows that the Sivers function is proportional
to the matrix elements involving hadron helicity flip. Because the
operator $\hat {\cal O}$ is chiral-even, it conserves the quark
helicities. To obtain the hadron helicity flip, we have to take
into account the nucleon's light-cone wave function with nonzero
quark orbital angular momentum, as in the previous two examples.

\begin{figure}[]
\begin{center}
\includegraphics[height=7.5cm]{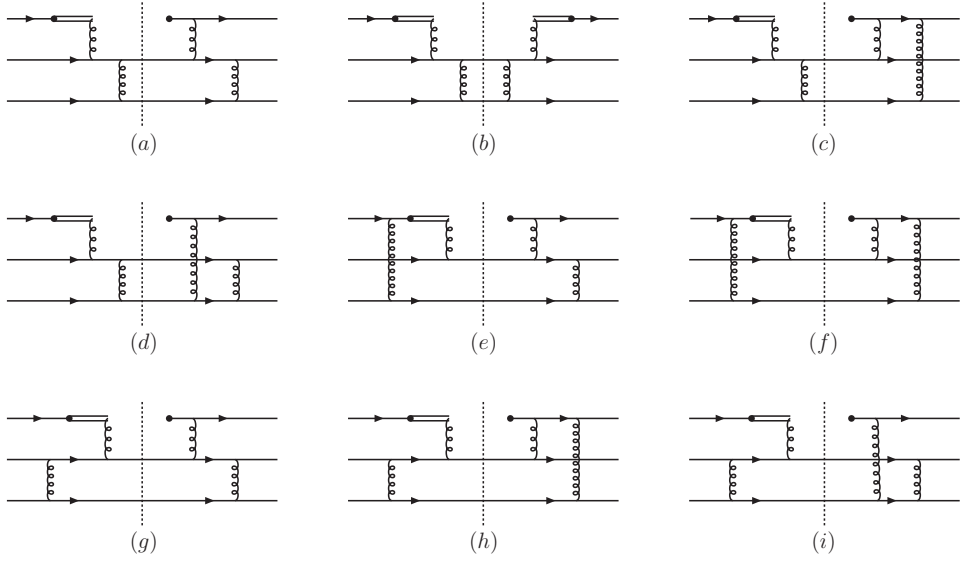}
\end{center}
\vskip -0.7cm \caption{\it Leading Feynman diagrams contributing
to the naive time-reversal-odd TMD quark distributions at
large-x.}
\end{figure}

Following the calculations in the last subsection for
$k_\perp$-odd distribution $\Delta q_T$, we find that the Sivers
function will depend on the following matrix element,
\begin{eqnarray}
{}_{\frac{1}{2}}\langle PS_{z\uparrow}|\hat{\cal
O}|PS_{z\downarrow}\rangle_{\frac{1}{2}}&+&{}_{-\frac{1}{2}}\langle
PS_{z\downarrow}|\hat{\cal
O}|PS_{z\uparrow}\rangle_{-\frac{1}{2}}\nonumber\\
+{}_{-\frac{1}{2}}\langle PS_{z\uparrow}|\hat{\cal
O}|PS_{z\downarrow}\rangle_{-\frac{1}{2}}&+&{}_{\frac{1}{2}}\langle
PS_{z\downarrow}|\hat{\cal O}|PS_{z\uparrow}\rangle_{\frac{1}{2}}\
. \label{ghalf}
\end{eqnarray}
Comparing with Eq.~(\ref{shalf}), we find that only the sign
changes in the above sum. This is because here we are probing the
unpolarized quark, and we have to sum up different quark helicity
contribution, while in Eq.~(\ref{shalf}) we are probing the
longitudinal polarized quark and different quark helicity will
contribute differently. From Eqs.~(\ref{e32},\ref{e33}), we find
that the final result for the above matrix element will be
\begin{equation}
 \langle PS_{x\uparrow}|\hat {\cal
O}|PS_{x\downarrow}\rangle +\langle PS_{x\downarrow}|\hat {\cal
O}|PS_{x\uparrow}\rangle\propto \int
\tilde\psi^{(1)}(y_i')\tilde\psi^{(3)}(y_i)\left(-ip_1^y\right)T_H(y_i;y_i';p_{i\perp})
+h.c.\ . \label{sm}
\end{equation}
From this equation, we find that in order to generate a nonzero
Sivers function, the hard scattering factor $T_H$ has to have an
imaginary part. In a partonic hard scattering amplitude, the only
imaginary part comes from the on-shell pole of some propagator. As
we showed in the above analysis all the propagators in Fig.~2 are
far off-shell except for the eikonal propagator from the gauge
link. Thus, in order to obtain nonzero contribution to the Sivers
function, we have to have eikonal propagator in the partonic
Feynman diagrams. We have shown all leading order diagrams
contribution to the Sivers function in Fig.~3, all of which have
at least one eikonal propagator. For example, the eikonal
propagator in Fig.~3(a) reads,
\begin{equation}
\frac{1}{n\cdot
(k-p_1)+i\epsilon}=P\frac{1}{x-y_1}-i\pi\delta(x-y_1)\ ,
\label{e49}
\end{equation}
where the first term is the principal value of the pole, and does
not contribute to the Sivers function. Only the second term
contribute to an imaginary part, which contains a delta function.
This delta function will affect the power counting for the various
factors in the evaluation of the matrix element of Eq.~(\ref{sm}).
This is because the delta function can be written as
$\delta(x-y_1)=\delta\left(y_2+y_3-(1-x)\right)$, which means that
the variables $y_2$ and $y_3$ are limited to be order of $(1-x)$,
i.e., $y_2\sim y_3\sim {\cal O}(1-x)$. All factors which depend on
$y_2$ and $y_3$ will have to be examined carefully to get the
right power counting results. For example, in Fig.~3(a), the
propagators at the left side of the cut will be affected by the
above constraints. One of the gluon propagator reads,
\begin{equation}
\frac{1}{(p_3-k_2)^2}\approx\frac{1}{-y_3\frac{\vec{k}_{2\perp}^2}{\beta(1-x)}}\approx
\frac{1}{\langle k_\perp^2\rangle} \ ,
\end{equation}
because $y_3\sim {\cal O}(1-x)$ and $\beta$ is order of unit.
Unlike the case studied in the above subsections, this propagator
does not lead to a suppression in $(1-x)$ for the Sivers function.
Similarly, another gluon propagator and the quark propagator at
the left side of the cut line are also finite at $x\to 1$.
However, all the propagators at the right hand side of the cut
line still scale as $(1-x)$, and the total four propagators there
will contribute to a suppression factor of $(1-x)^4$.

Another consequence of this delta function is that the intrinsic
$p_\perp$ expansion in the hard part has no additional suppression
in $(1-x)$, which is very different from what we have in the last
subsection for the $k_\perp$-odd but naive time-reversal-even
quark distributions. For example, in one of the above propagators,
we can keep the intrinsic transverse momentum dependence and
expand it up to the linear term,
\begin{equation}
\frac{1}{(p_3-k_2)^2}\approx\frac{1}{-2k_2\cdot p_3}
=\frac{1}{-\frac{y_3}{\beta(1-x)}k_{2\perp}^2-2k_{2\perp}\cdot
p_{3\perp}}\approx -\frac{1}{\langle k_\perp^2\rangle
}\left(1-\frac{k_{2\perp}\cdot p_{3\perp}}{\langle
k_\perp^2\rangle }\right) \ .
\end{equation}
The last equation of the above comes from the fact that $y_3\sim
{o}(1-x)$. After combining this expansion with the light-cone wave
function, we get $ \int p_{1\perp}^y k_{2\perp}\cdot p_{3\perp}
\tilde\psi^{(3)}\propto k_\perp^y y_3\Phi_4(y_i)~ {\rm or}~
y_2\Psi_4(y_i)
 $,  where the $k_\perp^y$ factor will cancel that in the denominator
in Eq.~(\ref{e46}).

The final step in this analysis will be to factor out $(1-x)$ from
the wave function integral. Because the variables $y_2$ and $y_3$
are constrained to be order of $(1-x)$ in the $x\to 1$ limit, we
have to examine the wave function integral where the end-point
behavior of the light-cone wave function will be important. For
example, one contribution in the above analysis is the integral of
$y_3\Phi_4(y_1,y_2,y_3)$. We can factor out the overall dependence
on $(1-x)$ from the integral,
\begin{eqnarray}
&&\int
dy_1dy_2dy_3\delta(1-y_1-y_2-y_3)\delta(y_1-x)y_3\Phi_4(y_1,y_2,y_3)
\nonumber\\
&&~~~=(1-x)^3\int d\zeta d\zeta' \delta
(1-\zeta-\zeta')\zeta'(1-x)\Phi_4(x,\zeta(1-x),\zeta'(1-x))/(1-x)^2
\ ,
\end{eqnarray}
where we have re-parameterized $y_2=\zeta (1-x)$ and
$y_3=\zeta'(1-x)$. The above integral depends on the end-point
behavior of the twist-four distribution amplitudes. As we showed
in the Appendix, the twist-four distribution amplitude behaviors
as $y_3 \Phi_4(y_1,y_2,y_3)\propto y_1 y_2 y_3$ at the end-point
region. From this, we find that
\begin{equation}
\lim_{x\to1}\frac{\zeta'(1-x)\Phi_4(x,\zeta(1-x),\zeta'(1-x))}{(1-x)^2}={\rm
finite} \ .
\end{equation}
Thus the wave function integral indeed contains a suppression
factor $(1-x)^3$. In addition, the Dirac matrix traces will also
result into a power dependence on $(1-x)$. This can be calculated
straightforwardly, and we find the Dirac traces from Fig.~3(a)
contribute to a power term as $(1-x)^{-2}$. By summarizing the
power counting results from the above analysis and also taking
into account the phase spaces integral factor $(1-x)^{-1}$ in
Eq.~(\ref{phase}), we find the Sivers function will have the
following power behavior,
\begin{equation}
q_T(x,k_\perp)|_{x\to 1}\propto (1-x)^4\ ,
\end{equation}
which is $(1-x)$ suppressed relative to the unpolarized quark
distribution. Similar calculations can be performed for all other
diagrams in Fig.~3, and they all contribute to a power behavior of
$(1-x)^4$ for the quark Sivers function.

The same analysis can be performed for another naive time-reversal
odd distribution, the so-called Boer-Mulders function $\delta q$,
which is defined as
\begin{equation}
\delta q=\frac{M_P}{2\epsilon^{ij}k_\perp^j}\int
        \frac{d\xi^-}{(2\pi)^3}e^{-ix\xi^-P^++i\xi_\perp\cdot k_\perp}\, \left\langle
P|\overline{\mit \Psi}_v(\xi)\gamma^+\gamma^i\gamma_5 {\mit
\Psi}_v(0)|P\right\rangle\ .
\end{equation}
If we choose $\gamma^i=\gamma^x$ in the above equation, the TMD
$\delta q$ will become,
\begin{equation}
\delta q(x,k_\perp)=\frac{M_P}{2k_\perp^y}\int
        \frac{d\xi^-}{(2\pi)^3}e^{-ix\xi^-P^++i\xi_\perp\cdot k_\perp} \langle
P|\hat {\cal O}_t|P\rangle\ ,
\end{equation}
where the operator $\hat{\cal O}_t$ follows the definition in the
above. Unlike the TMD quark distribution $\delta q_L$, in the
above definition the proton is unpolarized. Thus the explicit
expression for $\delta q$ in the proton helicity states is
\begin{equation}
\delta q(x,k_\perp)=\frac{M_P}{2k_\perp^x}\int
        \frac{d\xi^-}{(2\pi)^3}e^{-ix\xi^-P^++i\xi_\perp\cdot k_\perp} \frac{1}{2}\left(\langle
PS_{z\uparrow}|\hat {\cal O}_t|PS_{z\uparrow}\rangle+\langle
PS_{z\downarrow}|\hat {\cal O}_t|PS_{z\downarrow}\rangle\right)\ .
\end{equation}
Comparing this expression to Eq.~(\ref{h1l}) in the last
subsection, we find that $\delta q$ depends on the sum of the two
matrix elements, whereas for $\delta q_L$ it is the difference. As
in the analysis for the Sivers function, we find that we need an
imaginary part from the hard part, and the same set of diagrams in
Fig.~3 contribute. The final result for its power counting will be
\begin{equation}
\delta q|_{x\to 1}\sim (1-x)^4 \ ,
\end{equation}
which is again one power of $(1-x)$ suppressed relative to the
unpolarized quark distribution.

\subsection{Comparison with the Power Counting for the  GPD $E$}

Summarizing the results in the last two subsections, we find that
the $k_\perp$-odd TMD quark distributions are suppressed by a
relative factor $(1-x)$ to the $k_\perp$-even ones (e.g., the
unpolarized quark distributions). As we mentioned in the
introduction, this can also be understood by the interpretation of
these TMD quark distributions in terms of the overlaps of the
light-cone wave functions of $L_z=0$ and $|L_z|=1$ Fock states.

As in the case of the Pauli form factor, the generalized parton
distribution (GPD) $E$ and the Sivers function $q_T$ involve the
overlap of initial- and final-state light-front-wave-functions
(LFWFS) which differ by one unit of orbital angular momentum
\cite{Brodsky:2000xy}. In contrast to the Sivers function which is
suppressed by one power of $(1-x)$ relative to the unpolarized
distribution, one finds the GPD $E$ falls as two-powers $(1-x)^2$
faster than the spin-conserving GPD $H$ at large-$x$ \cite{yuan}.
The power of $(1-x)^n$ thus differs when we compare the $E$ GPD
arising in spin-flip deeply virtual Compton scattering (DVCS)
$\gamma^* p_\downarrow \to \gamma p_\uparrow$ and the Sivers
function $q_T$ arising in polarized electroproduction $\gamma^*
p_\updownarrow \to \pi X.$  In the following, we will briefly
comment why this happens.

It  is useful to use the symmetric light-front (LF) frame where
the transverse momenta of the initial and final state proton
momentum changes from $\vec {p}^{~\bf initial}_\perp =( \vec
p_\perp - {1\over 2} \Delta_\perp)$ to $\vec {p}^{~\bf final} =
(\vec p_\perp + {1\over 2} \Delta_\perp).$ The struck quark in
DVCS is evaluated at $\vec k_\perp + {1\over 2}(1-x)\vec
\Delta_\perp$ in the final-state LFWF and $\vec k_\perp  - {1\over
2}(1-x)\vec \Delta_\perp$ in the initial-state LFWF,  as in the
Drell-Yan-West (DYW) formula for current matrix elements
\cite{dyw}.

The  $E$ GPD  requires evaluating the spin-flip deeply virtual
Compton amplitude which is linear in the transverse momentum
transfer to the proton $\vec \Delta_\perp$. This kinematic factor
arises from the extra angular momentum of the initial- or
final-state LFWF with argument $ \pm {1\over 2} (1-x) \vec
\Delta_\perp$. In addition, the orbital angular momentum dynamics
of the LFWF introduces a factor of $(1-x)$. Thus $E \sim (1-x)^2
H$ as $x \to 1 .$

In contrast, when we evaluate the Sivers SSA for SIDIS $\gamma^* p
\to \pi  p^\prime$, the dynamics of the orbital angular momentum
in the LFWF gets expressed as the transverse momentum  $\vec
p_{\pi \perp}$ of the produced pion, not the change in the
transverse momentum $\vec \Delta_\perp$ of the  proton.  Thus the
second factor of $(1-x)$ does not appear in the Sivers function.
We thus have the power counting rule: $E \sim  (1-x) q_T
\sim (1-x)^2 H$ as $x \to 1 .$

\subsection{Power counting for the integrated quark distributions
at leading and higher-twist}

From the power counting results for the TMD quark distributions in
the last subsections, we can further derive the power counting
rule for the integrated quark distributions at large-x when
integrating over the transverse momentum. For example, the
integrated unpolarized quark distribution can be written as
\begin{equation}
q(x)=\int d^2\vec{k}_\perp q(x,k_\perp) \ .
\end{equation}
Similar equations also hold for the longitudinal polarized quark
distribution and transversity quark distribution. From the power
counting of these relevant TMD quark distributions, we can
immediately see that the integrated quark distributions have the
following power counting rule at $x\to 1$,
\begin{equation}
q(x)\sim (1-x)^3,~~~\Delta q_L(x)\sim (1-x)^3,~~~\delta q_T(x)\sim
(1-x)^3 \ .
\end{equation}
To obtain the above power counting results for the integrated
quark distributions, we have assumed that the $k_\perp$-integral
decouples from the $x-$distributions of the partons
\cite{brolep,gunion}. Although the upper limit of the $k_\perp$
integral might depend on $(1-x)$, the bulk of this integration
comes from the lower bound, which will not affect the $(1-x)$
power counting for the integrated parton distributions \cite{fac}.
The latest comparison of the above power counting predictions with
experiment can be found in \cite{zheng}.

The $k_\perp$-moment of the $k_\perp$-odd TMD quark distributions
are related to the twist-three parton distributions. For example,
the twist-three parton distribution $g_T(x)$ is related to the TMD
quark distribution $\Delta g_T$ \cite{jmy03},
\begin{equation}
g_T(x)=\frac{1}{2xM^2}\int d^2\vec{k}_\perp \vec{k}_\perp^2\Delta
q_T(x,k_\perp) \ ,
\end{equation}
and for $h_L(x)$,
\begin{equation}
h_L(x)=\frac{-1}{xM^2}\int d^2\vec{k}_\perp \vec{k}_\perp^2\delta
q_L(x,k_\perp) \ .
\end{equation}
Of course, caution has to be taken when we apply the above
equations \cite{jmy03}. From the power counting rule for the
relevant TMD quark distributions, we find the following power
behavior for these two twist-three parton distributions,
\begin{equation}
g_T(x)\sim (1-x)^4,~~~h_L(x)\sim (1-x)^4 \ .
\end{equation}
$k_\perp$-moment of the naive-time-reversal-odd TMD quark
distributions are also related to the twist-three parton
distributions, which have been shown in literature \cite{bmp}, for
example,
\begin{equation}
T_F(x)=\frac{1}{M_P} \int d^2\vec{k}_\perp \vec{k}_\perp^2
q_T(x,k_\perp) \ ,
\end{equation}
where $T_F$ is the so-called Qiu-Sterman matrix element \cite{qs},
and is responsible to the SSA for inclusive hadron production in
hadronic collisions. Similarly, the $k_\perp$-moment of $\delta q$
corresponds,
\begin{equation}
T_F^{(\sigma)}(x)=\frac{1}{M_P} \int d^2\vec{k}_\perp
\vec{k}_\perp^2 \delta q(x,k_\perp) \ ,
\end{equation}
where $T_F^{(\sigma)}$ is defined as
\begin{equation}
T_F^{(\sigma)}(x_1,x_2)=
\int\frac{d\zeta^-d\eta^-}{8\pi}e^{ix_1P^+\eta^-}e^{i(x_2-x_1)P^+\zeta^-}
\left\langle P|\overline\psi(0)\sigma
^{+\alpha}gF^{+\alpha}(\zeta^-)\psi(\eta^-)|P\right\rangle \ ,
\end{equation}
and $T_F^{(\sigma)}(x)\equiv T_F^{(\sigma)}(x,x)$. From the power
counting rule of the relevant naive-time-reversal-odd TMD quark
distributions, we find the power counting rule for these two
twist-three parton distributions,
\begin{equation}
T_F(x)\sim (1-x)^4,~~~T_F^{(\sigma)}(x)\sim (1-x)^4 \ ,
\end{equation}
which are $(1-x)$ suppressed relative to the unpolarized quark
distribution.

\subsection{TMD quark distributions in Pion}

\begin{figure}[]
\begin{center}
\includegraphics[height=5.0cm]{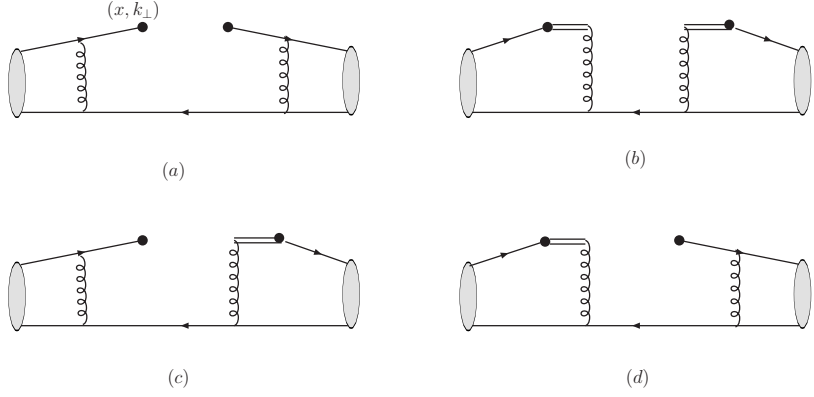}
\end{center}
\vskip -0.7cm \caption{\it Feynman diagrams contribution to the
transverse-momentum-dependent quark distribution in Pion at
large-x.}
\end{figure}

A similar analysis can be carried out for the TMD quark
distributions of pion. Because the pion is a spin-0 particle,
there are only two leading-order TMD quark distributions: the
unpolarized quark distribution $q_\pi(x,k_\perp)$ and the
Boer-Mulders function $\delta q_\pi(x,k_\perp)$. At large-x, their
dependence on $x$ can be calculated from the diagrams shown in
Fig.~4. The unpolarized quark distribution at large-x will have
contributions from all of the four diagrams,
\begin{equation}
u_\pi(x,k_\perp)=\frac{f_\pi^2}{(k_\perp^2)^2}(1-x)^2\alpha_s^2C_F\int
\frac{dz_1}{z_1}\frac{dz_2}{z_2}\Phi_\pi(z_1)\Phi_\pi(z_2){\cal
T}_H(z_1,z_2) \ ,
\end{equation}
where $f_\pi$ is the decay constant of pion, and $\Phi_\pi(z)$ is
the leading-twist quark distribution amplitude. Obviously, the
quark distribution has $(1-x)^2$ power behavior at large-x. This
is consistent with the Gribov-Liptov relation \cite{gribov}.

For the Boer-Mulders function, because it is
naive-time-reversal-odd, we have to take into account the
interference between $L_z=0$ and $|L_z|=1$ Fock states of pion
light-cone wave functions, and also the gauge link is important to
obtain a phase difference. Following the same analysis in the
previous subsections, we find that the Boer-Mulders function of
pion has the same power counting result as the unpolarized quark
distribution,
\begin{equation}
\delta q_\pi(x,k_\perp)\sim (1-x)^2 \ .
\end{equation}
These two distribution functions having the same power behavior at
large-x, is not because the Boer-Mulders function of pion gets
enhancement, but because the unpolarized quark distribution of
pion is suppressed by one power of $(1-x)$ compared to the usual
power counting results for parton distributions of hadrons
\cite{jackson,gunion}.

\section{conclusion}

In this paper, we have performed a perturbative analysis of the
transverse momentum dependent quark distributions at large x. A
generalized power counting rule has been derived for the leading
order TMD quark distributions, and we have found that the
$k_\perp$-even distributions all scale as $(1-x)^3$, whereas the
$k_\perp$-odd ones as $(1-x)^4$, including the naive
time-reversal-even and -odd distributions. In particular, we have
shown that the quark Sivers function has power behavior of
$(1-x)^4$, which is $(1-x)$ suppressed relative to the unpolarized
quark distribution. For the TMD quark distributions of pion, we
find that the Boer-Mulders function has the same power behavior as
the unpolarized quark distribution, scaling as $(1-x)^2$ in the
limit. These results provide important guidelines for the
parameterizations of the transverse momentum dependent parton
distributions and the quark-gluon correlation functions in the
phenomenological studies.

In our analysis we have not included the effects of perturbative
QCD evolution. In fact, the evolution  of parton distributions at
large $x$ with photon virtuality $Q^2$ is suppressed compared to
the usual DGLAP evolution \cite{brolep} because the struck quark
is a bound constituent of the target hadron.  In particular, in
the limit of $x\to 1$, the virtuality of the struck quark becomes
highly space-like, and evolution is effectively quenched. Thus the
power counting of structure functions at large $x$ is not affected
by evolution \cite{brolep},  allowing duality with the power-law
falloff of exclusive channels at fixed $W^2$. Another important
point has to be kept in mind is the large logarithms associated
with the parton distributions in the $x\to 1$ limit, in terms of
$\alpha_s^n{\rm log}^m(1/(1-x))$ for $m\le 2n$
\cite{brolep,mul,{Korchemsky:1988si},{Sterman:1986aj}}. All these
effects will of course introduce additional theoretical
uncertainties when we apply the power counting rule to the parton
distributions at large-x.

\section*{Acknowledgments}
We thank Xiangdong Ji, Jianwei Qiu, and Werner Vogelsang for their
comments. S.J.B. is supported by the Department of Energy under
contract number DE--AC02--76SF00515, F.~Y. is grateful to RIKEN,
Brookhaven National Laboratory and the U.S. Department of Energy
(contract number DE-AC02-98CH10886) for providing the facilities
essential for the completion of their work. Finally, F.Y. thanks
Beijing Institute of Modern Physics, Institute of Theoretical
Physics, Beijing University, for its warm host when part of this
work was finished.

\appendix
\section{Light-cone wave functions and distribution amplitudes of
nucleon}

In this Appendix we list the light-cone wave functions for the
three-quark Fock states of nucleon as reference \cite{jmy03},
\begin{eqnarray}
  |PS_{z\uparrow}\rangle &=& \int d[1]d[2]d[3] \left\{\tilde \psi^{(1)}(1,2,3)
          u^{\dagger}_{\uparrow}(1)
             \left(u^{\dagger}_{\downarrow}(2)d^{\dagger}_{\uparrow}(3)
            -d^{\dagger}_{\downarrow}(2)u^{\dagger}_{\uparrow}(3)\right)
         |0\rangle \nonumber\right.\\
         &&\left.+\left((p_1^x+ip_1^y)
         \tilde \psi^{(3)}(1,2,3)
         + (p_2^x+ip_2^y) \tilde
         \psi^{(4)}(1,2,3)\right)\right.\nonumber\\
         &&\left. \left( u^{\dagger}_{\uparrow}(1)
            u^{\dagger}_{\downarrow}(2)d^{\dagger}_{\downarrow}(3)
            -d^{\dagger}_{\uparrow}(1)u^{\dagger}_{\downarrow}(2)
             u^{\dagger}_{\downarrow}(3)\right)
         |0\rangle \right\}\ ,\nonumber\\
  |PS_{z\downarrow}\rangle &=& \int d[1]d[2]d[3] \left\{-\tilde \psi^{(1)}(1,2,3)
          u^{\dagger}_{\downarrow}(1)
             \left(u^{\dagger}_{\uparrow}(2)d^{\dagger}_{\downarrow}(3)
            -d^{\dagger}_{\uparrow}(2)u^{\dagger}_{\downarrow}(3)\right)
         |0\rangle \nonumber\right.\\
         &&\left.+\left((p_1^x-ip_1^y)
         \tilde \psi^{(3)}(1,2,3)
         + (p_2^x-ip_2^y) \tilde
         \psi^{(4)}(1,2,3)\right)\right.\nonumber\\
         &&\left.
         \left( u^{\dagger}_{\downarrow}(1)
            u^{\dagger}_{\uparrow}(2)d^{\dagger}_{\uparrow}(3)
            -d^{\dagger}_{\downarrow}(1)u^{\dagger}_{\uparrow}(2)
             u^{\dagger}_{\uparrow}(3)\right)
         |0\rangle \right\}\ ,
\label{w1}
\end{eqnarray}
where the argument $i$ is the shorthand for quark momentum
variables $y_i$ and $p_{i\perp}$, and the measure for the quark
momentum integrations is
\begin{eqnarray}
    d[1]d[2]d[3] &=& \sqrt{2}\frac{dy_1dy_2dy_3}{\sqrt{2y_1 2y_2 2y_3}}
                  \frac{d^2\vec{p}_{1\perp}d^2
             \vec{p}_{2\perp}d^2\vec{p}_{3\perp}}{(2\pi)^9}
       \nonumber \\
                  && \times 2\pi\delta(1-y_1-y_2-y_3)(2\pi)^2\delta^{(2)}
                   (\vec{p}_{1\perp}+\vec{p}_{2\perp}+\vec{p}_{3\perp}) \ .
\end{eqnarray}
$\tilde\psi^{(1,3,4)}$ are the light-cone wave function amplitudes
for the three quark Fock state expansion of nucleon.
$\tilde\psi^{(1)}$ corresponds to the $L_z=0$ Fock state
component, and $\tilde\psi^{(3,4)}$ for $|L_z|=1$ ones. These
light-cone wave functions were used in our analysis for the
large-x quark distributions.

In order to get Eq.~(\ref{e5}), the light-cone wave functions have
to be converted into the quark distribution amplitudes
\cite{brolep}. For example, we can integrate out the transverse
momentum in the leading Fock state light-cone wave function, and
define the twist-three amplitude,
\begin{equation}
    \Phi_3(y_i) = 2\sqrt{6}\int  \frac{d^2\vec{p}_{1\perp}d^2
             \vec{p}_{2\perp}d^2\vec{p}_{3\perp}}{(2\pi)^6}
      \delta^{(2)}(\vec{p}_{1\perp}+\vec{p}_{2\perp}+\vec{p}_{3\perp})
       \tilde \psi^{(1)}(1,2,3) \ .
\end{equation}
For $|L_z|=1$ states, we have to keep linear term in the $p_\perp$
expansion of the hard factor, and combine them with the light-cone
wave function, which will lead to the twist-four distribution
amplitudes of the nucleon \cite{braun,f2},
\begin{eqnarray}
    \Psi_4(y_1,y_2,y_3) &=&  \frac{2\sqrt{6}}{y_2M}\int  \frac{d^2\vec{p}_{1\perp}d^2
             \vec{p}_{2\perp}d^2\vec{p}_{3\perp}}{(2\pi)^6}
      \delta^{(2)}(\vec{p}_{1\perp}+\vec{p}_{2\perp}+\vec{p}_{3\perp}) \nonumber \\ &&
  \times   \vec{p}_{2\perp}\cdot \left[   \vec{p}_{1\perp} \tilde \psi^{(3)}(1,2,3)
 + \vec{p}_{2\perp} \tilde \psi^{(4)}(1,2,3)\right] \ .
      \nonumber \\
    \Phi_4(y_2,y_1,y_3) &=&  \frac{2\sqrt{6}}{y_3M}\int  \frac{d^2\vec{p}_{1\perp}d^2
             \vec{p}_{2\perp}d^2\vec{p}_{3\perp}}{(2\pi)^6}
      \delta^{(2)}(\vec{p}_{1\perp}+\vec{p}_{2\perp}+\vec{p}_{3\perp}) \nonumber \\ &&
  \times  \vec{p}_{3\perp}\cdot \left[   \vec{p}_{1\perp} \tilde \psi^{(3)}(1,2,3)
 +  \vec{p}_{2\perp} \tilde \psi^{(4)}(1,2,3)\right] \ .
\end{eqnarray}
The explicit expressions for these distribution amplitudes are not
necessary for the power-counting analysis. However, the end-point
behavior at $y_i\to 1$ is needed for the power counting of the
naive time-reversal-odd TMD quark distributions. We note that in
the end-point region, these distribution amplitudes have the
following behaviors: $\Phi_3(y_i)\propto y_1y_2y_3$,
$y_2\Psi_4(y_i)\propto y_1y_2y_3$, and $y_3\Phi_4(y_i)\propto
y_1y_2y_3$ \cite{braun}. From this, we immediately find that the
end-point behavior of the $p_\perp$-moment of the light-cone wave
functions. For example,
\begin{eqnarray}
\int  {d^2\vec{p}_{1\perp}d^2
             \vec{p}_{2\perp}d^2\vec{p}_{3\perp}}
      \delta^{(2)}(\vec{p}_{1\perp}+\vec{p}_{2\perp}+\vec{p}_{3\perp})
      (\vec{p}_{i\perp}\cdot \vec{p}_{j\perp}) \tilde
      \psi^{(3,4)}(y_1,y_2,y_3)|_{\rm end~point}\sim y_1y_2y_3  \ ,
\end{eqnarray}
where $i,j=1,2,3$. These properties have been used in our
analysis.

\end{document}